\documentclass{iau} 
\usepackage{graphicx,natbib}

\title[Ly$\alpha$ emitting galaxies at high redshifts] 
{The VIMOS Ultra Deep Survey: Ly$\alpha$ Emission  and Stellar
  Populations of Star-Forming Galaxies at 2\,$<$\,$z$\,$<$\,6}

\author[Nimish P. Hathi \& the VUDS team]   
{Nimish P. Hathi$^1$,
 Olivier Le F\`evre$^1$, and the VUDS team\footnote{{\scriptsize P.~Cassata,
B.~Garilli, V.~Le Brun, B.~C.~Lemaux, D.~Maccagni, L.~Pentericci,
L. A. M.~Tasca, R.~Thomas, E.~Vanzella, G.~Zamorani, E.~Zucca,
R.~Amorin, S.~Bardelli, L.~P.~Cassar\`a, M.~Castellano, A.~Cimatti,
O.~Cucciati, A.~Durkalec, A.~Fontana, M.~Giavalisco, A.~Grazian,
O.~Ilbert, S.~Paltani, B.~Ribeiro, D. Schaerer, M.~Scodeggio,
V.~Sommariva, M.~Talia, L.~Tresse, D.~Vergani, P.~Capak, S.~Charlot,
T.~Contini, S.~de la Torre, J.~Dunlop, S.~Fotopoulou, A.~Koekemoer,
C.~L\'opez-Sanjuan, Y.~Mellier, J.~Pforr, M.~Salvato, N.~Scoville, Y.~Taniguchi, 
and P.W. Wang.}
}}

\affiliation{$^1$Aix Marseille Universit\'e, CNRS, LAM (Laboratoire d'Astrophysique de Marseille) UMR 7326, 13388, Marseille, France \\ email: {\tt nimish.hathi@lam.fr}} 

\pubyear{2015}
\volume{319}  
\jname{Galaxies at High Redshift and Their Evolution over Cosmic Time}
\editors{Sugata Kaviraj, \& Henry C. Ferguson, eds.}

\newcommand\lya{Ly$\alpha$}
\newcommand{\figref}[1]{Figure~\ref{#1}}
\newcommand\sfgn{SFG$_{\rm N}$}
\newcommand\sfgl{SFG$_{\rm L}$}

\begin{document}

\maketitle

\begin{abstract}
  The extensive ground-based spectroscopy campaign from the VIMOS
  Ultra-Deep Survey (VUDS), and the deep multi-wavelength photometry
  in three very well observed extragalactic fields (ECDFS, COSMOS,
  VVDS), allow us to investigate physical properties of a large sample
  ($\sim$4000 galaxies) of spectroscopically confirmed faint
  ($i_{AB}$\,$\lesssim$\,25~mag) SFGs, with and without \lya\ in
  emission, at $z$\,$\sim$\,2--6.  The fraction of \lya\ emitters
  (LAEs; equivalent width (EW)\,$\geq$\,20\AA) increases from
  $\sim$10\% at $z$\,$\sim$\,2 to $\sim$40\% at $z$\,$\sim$\,5--6,
  which is consistent with previous studies that employ higher \lya\
  EW cut. This increase in the LAE fraction could be, in part, due to
  a decrease in the dust content of galaxies as redshift
  increases. When we compare best-fit SED estimated stellar parameters
  for LAEs and non-LAEs, we find that E$_{\rm s}$(B-V) is smaller for
  LAEs at all redshifts and the difference in the median
  E$_{\rm s}$(B-V) between LAEs and non-LAEs increases as redshift
  increases, from 0.05 at $z$\,$\sim$\,2 to 0.1 at $z$\,$\sim$\,3.5 to
  0.2 at $z$\,$\sim$\,5.  For the luminosities probed here
  ($\sim$L$^*$), we find that star formation rates (SFRs) and stellar
  masses of galaxies, with and without Ly$\alpha$ in emission, show
  small differences such that, LAEs have lower SFRs and stellar masses
  compared to non-LAEs.  This result could be a direct consequence of
  the sample selection.  Our sample of LAEs  are selected
  based on their continuum magnitudes and they probe higher continuum
  luminosities compared to narrow-band/emission line selected
  LAEs. Based on our results, it is important to note that all LAEs
  are not universally similar and their properties are strongly
  dependent on the sample selection, and/or continuum luminosities.

\end{abstract}

\firstsection 
\section{Introduction}

In recent years, the unprecedented increase in the sensitivity of the
space-based as well as the ground-based observations has
revolutionized our understanding of high redshift ($z$\,$\gtrsim$\,2)
galaxies \citep[e.g.,][]{fink15,bouw15,elli13,hath10}.  This large
reservoir of star-forming galaxies (SFGs) has tremendous implications
on our understanding of the process of galaxy formation and evolution.
Lyman alpha (Ly$\alpha$) is typically the strongest UV emission line
in SFGs and a crucial spectroscopic signature to confirm high redshift
galaxies selected based on their colors.  The first studies of
Ly$\alpha$ emitters (LAEs) predicted that they could represent the
first galaxies in formation \citep[e.g.,][]{part67}.  Although
originally predicted to be extremely young, recent studies have shown
that LAEs have a variety of
ages, from 1 Myr to 1 Gyr \citep[e.g.,][]{gawi06,fink07,lai08,korn10},
range in dust extinction \citep[e.g.,][]{pirz07,fink09}, and a wide
range in stellar masses \citep[e.g.,][]{shap03,erb06,pent07,hath15}.
Such a large diversity in physical properties of LAEs implies that
these are not galaxies undergoing their first burst of star formation.
It is also puzzling that some LAEs show high dust content as
Ly$\alpha$ photons cannot easily escape from dusty galaxies because they
are resonantly scattered by neutral hydrogen.  These results, which
are based on both narrow-band (NB) as well as broad-band selection,
show a wide range of stellar properties for LAEs which contradicts
early predictions of LAEs as young, first galaxies.  In the era of
large surveys, it is now possible to study statistically significant
sample of these galaxies at all redshifts and get better insight into
the physical nature of LAEs, which has important implications on our
understanding of evolutionary properties of galaxies and the state of
intergalactic medium (IGM) in the early universe.

\vspace{-0.15in}
\section{Observations and Sample Properties}

The VUDS observations were done using the low-resolution multi-slit
mode of VIMOS on the VLT. A total of 15 VIMOS pointings ($\sim$224
arcmin$^2$ each, $\sim$1 deg$^2$ total) were observed covering the
full wavelength range from 3650\AA\ to 9350\AA\ in three deep survey
fields (ECDFS, VVDS-02h, COSMOS), which has extensive multi-wavelength
data. The primary selection criterion for galaxies in the VUDS program
was photometric redshifts. Therefore, the targets for the VUDS program
include a representative sample of all SFGs at a particular redshift
within a given magnitude limit ($i_{AB}$\,$\lesssim$\,25~mag, with
some galaxies as faint as $i_{AB}$\,$\sim$\,27~mag). A detailed
discussion about these observations, data reduction process, target
selection, reliability of the redshift measurements and corresponding
quality flags is presented in \citet{lefe15}.

\begin{figure}[t]
\minipage{0.33\textwidth}
\includegraphics[width=0.99\textwidth]{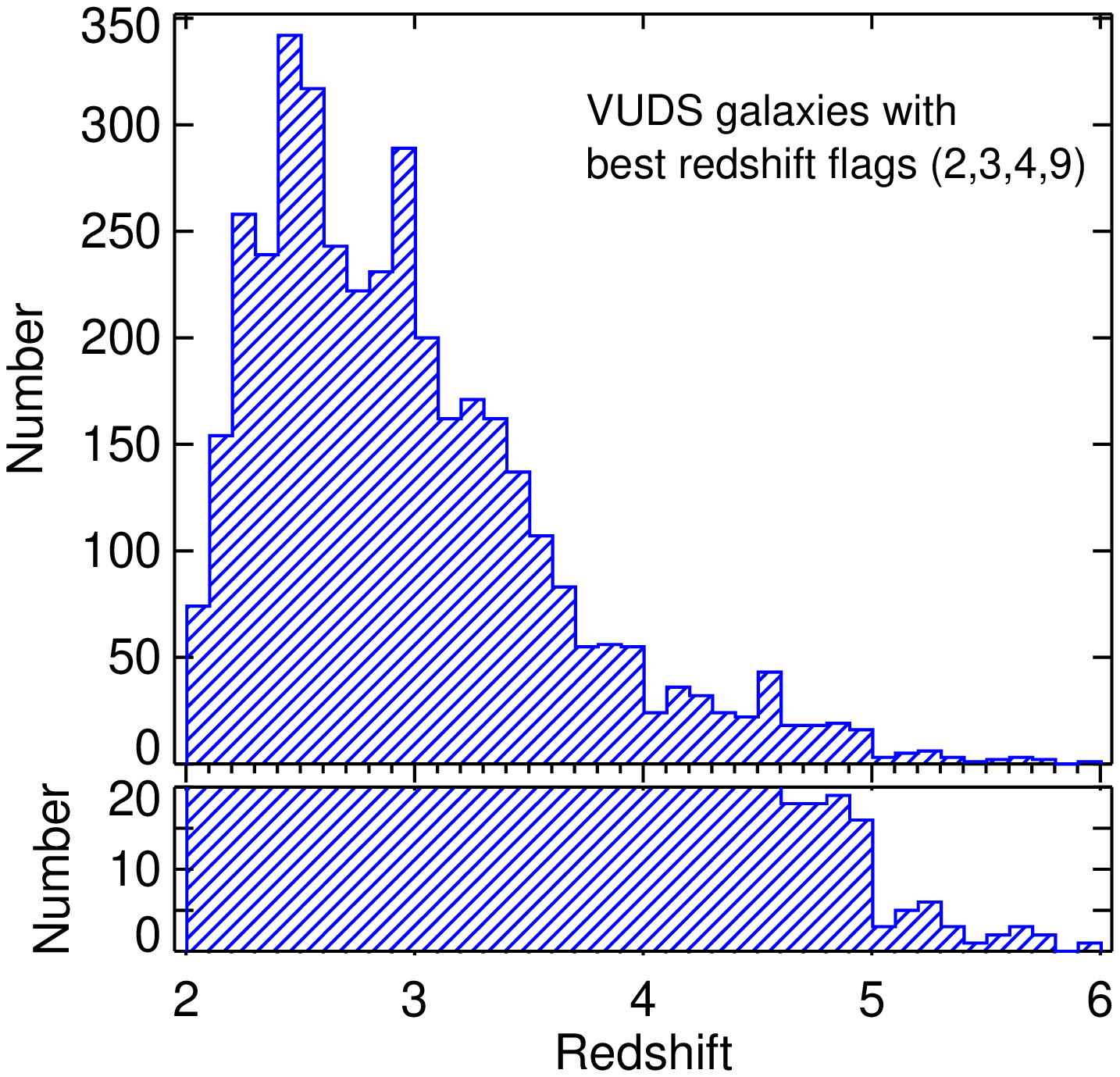} 
\endminipage\hfill
\minipage{0.33\textwidth}
\includegraphics[width=0.99\textwidth]{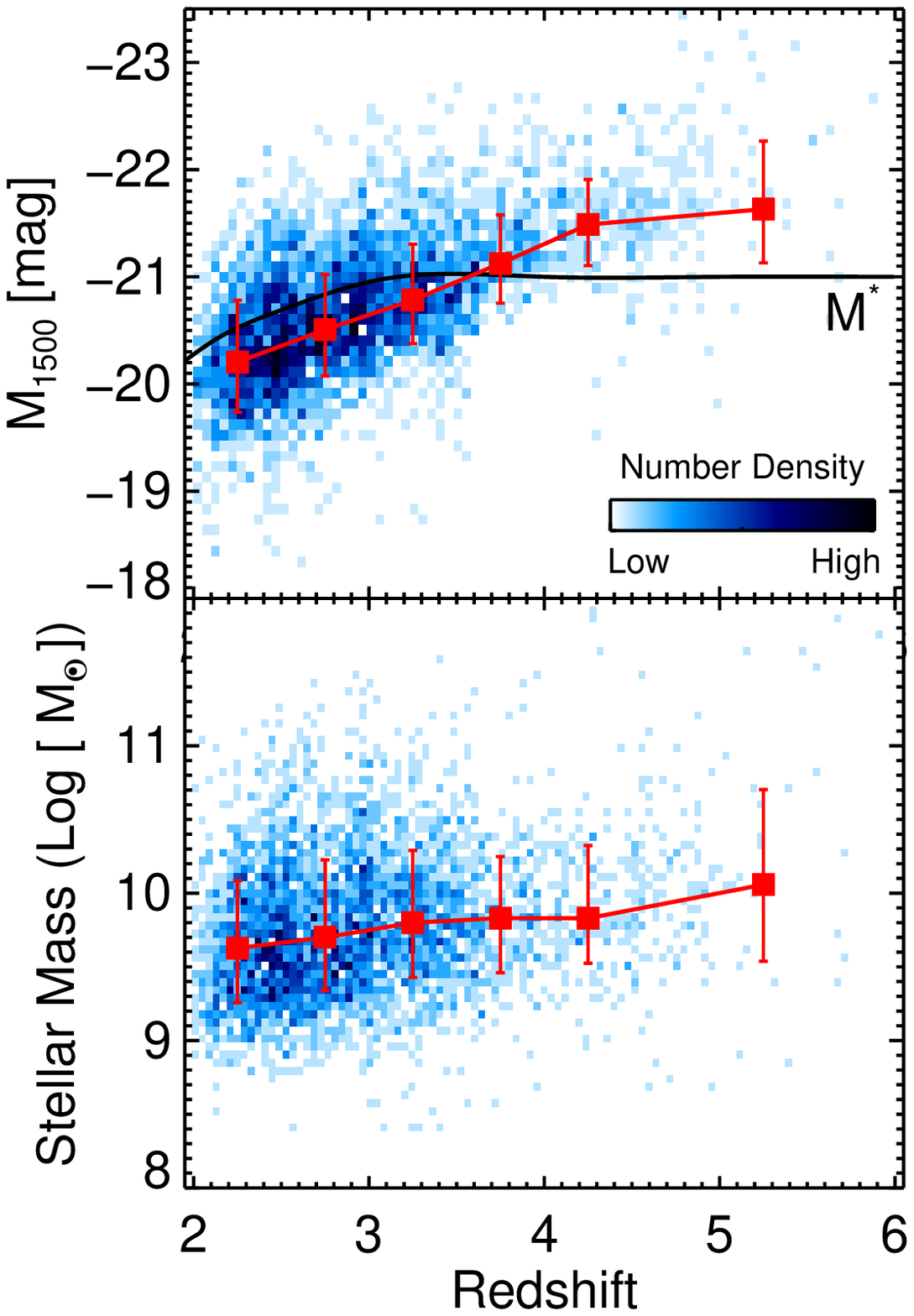}
\endminipage\hfill
\minipage{0.33\textwidth}
\includegraphics[width=0.99\textwidth]{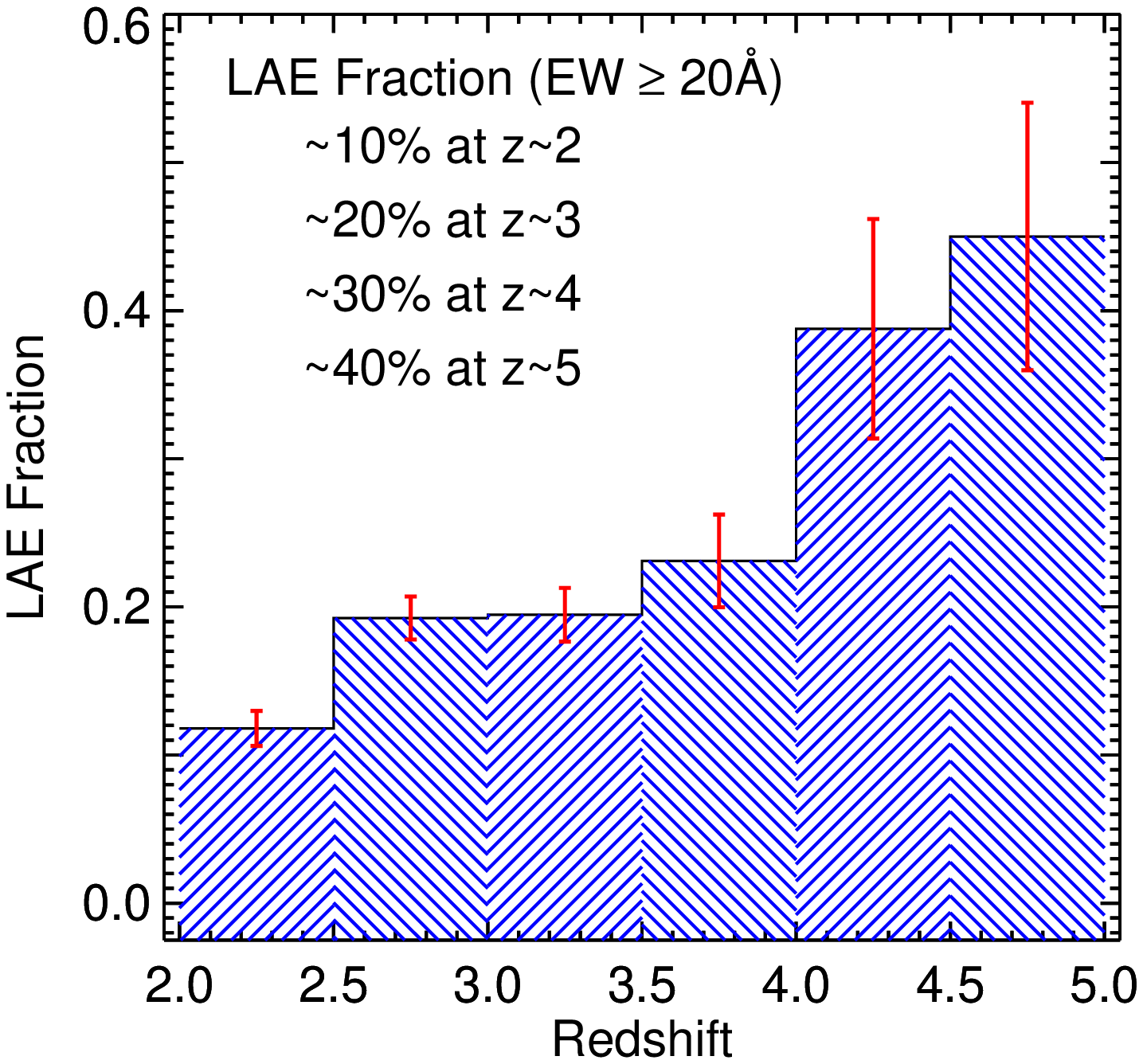}
\endminipage
\caption{[Left] The VUDS spectroscopic redshift distribution of SFGs
  at $z$\,$\sim$\,2--6 in our sample, which includes only the
  redshifts with best quality flags.  [Top-middle] The UV absolute
  magnitude, M$_{\rm 1500}$, as a function of redshift. The density of
  points (in both middle panels) is color-coded as shown in the
  color-bar. The median (red squares) M$_{\rm 1500}$ values with 16
  and 84 percentiles (error bars) are shown for each redshift bin. The
  black line shows the evolution of the characteristic magnitude,
  M$^*$, based on the values from \citet{hath10} and
  \citet{fink15}. [Bottom-middle] The distribution of stellar mass as
  a function of redshift. The median (red squares) stellar mass values
  with 16 and 84 percentiles (error bars) are shown for each redshift
  bin. [Right] The LAE fraction as a function of redshift. Here, LAEs
  are defined as galaxies with rest-frame \lya\ EW\,$\geq$\,20\AA,
  where positive EWs indicate \lya\ in emission. }
   \label{sample}
\end{figure}

We select all VUDS objects between $z$\,=\,2 and $z$\,=\,6, keeping
only the best reliability flags (2,3,4,9) --- which gives very high
probability (75-85\%, 95-100\%, 100\%, 80\%, respectively; see
\citealt{lefe15} for details) for these redshifts to be correct. The
redshift distribution of the sample is shown in the left panel of
\figref{sample}. Our sample of SFGs has only little contamination
from AGN identified based on their X-ray emission and IRAC colors
($\sim$2--3\%).
The \texttt{Le PHARE} software package \citep{ilbe06} was used to fit
the broad-band observed spectral energy distributions (SEDs) with
synthetic  stellar population models.
A detailed discussion about the SED fitting process is presented in
\citet{hath15}. From the best-fit model, we estimate stellar mass,
dust extinction E$_{\rm s}$(B-V), star-formation rate (SFR), and stellar
age for each galaxy.

The middle panel of \figref{sample} shows UV (1500\AA) absolute
magnitudes (M$_{\rm 1500}$) and stellar masses as a function of
redshift for the VUDS SFG sample.  We are probing UV continuum
luminosities around L$^*$ (or brighter) at these redshifts, and
similar median stellar masses (within error bars) at all redshifts.

The \lya\ equivalent widths (EWs) for VUDS SFGs were measured as
described in \citet{cass15} and \citet{hath15}.  We divide the SFG
population into three sub-groups based on their \lya\ EW. The galaxies
that show no \lya\ in emission (EW\,$\le$\,0\AA) are defined as \sfgn,
while the galaxies with \lya\ in emission, irrespective of its
strength (EW\,$>$\,0\AA), are defined as \sfgl. The third group is for
strong \lya\ emitters (EW\,$\ge$\,20\AA) called LAEs.  The fraction of
LAEs in SFGs increases from $\sim$10\% at $z$\,$\sim$\,2 to $\sim$40\%
at $z$\,$\sim$\,5--6 as shown in the right panel of \figref{sample}.
This result is in agreement with the general scenario that the
fraction of LAEs in SFGs increases as redshift increases reaching
$\sim$30-40\% at $z$\,$\simeq$\,6 \citep[e.g.,][]{star10,cass15}.

\begin{figure}[t]
\begin{center}
{\includegraphics[width=0.32\textwidth]{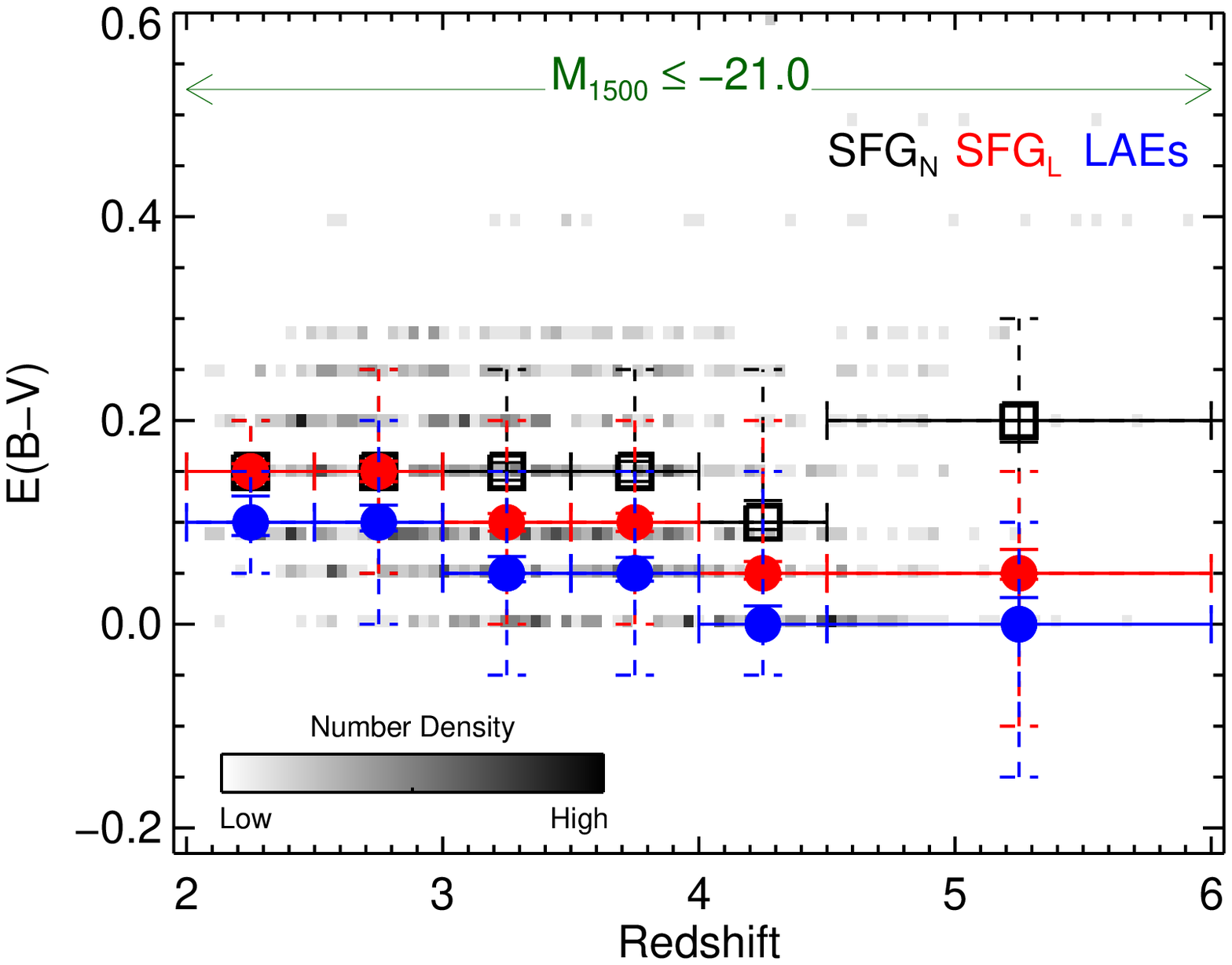} 
\includegraphics[width=0.32\textwidth]{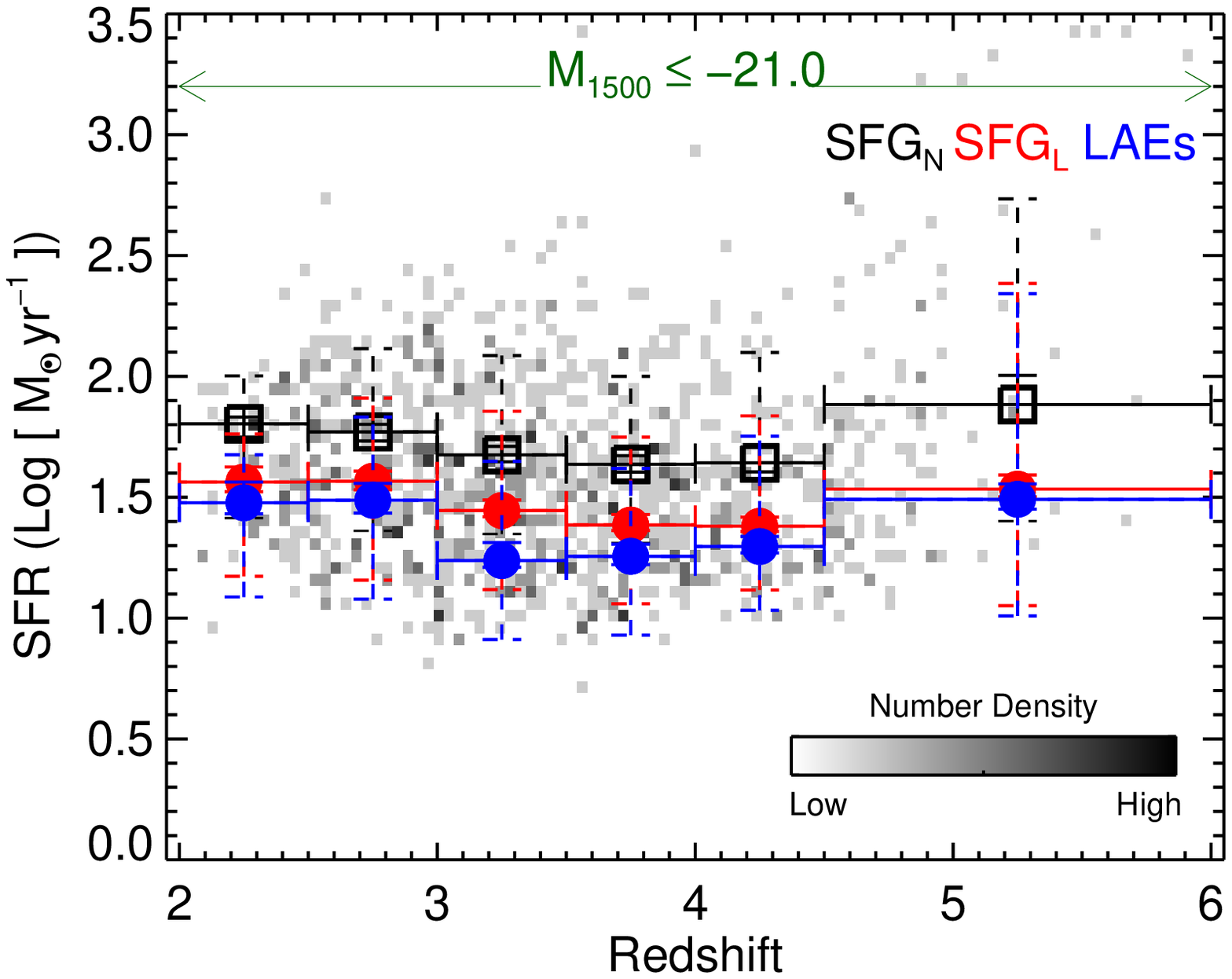}
\includegraphics[width=0.32\textwidth]{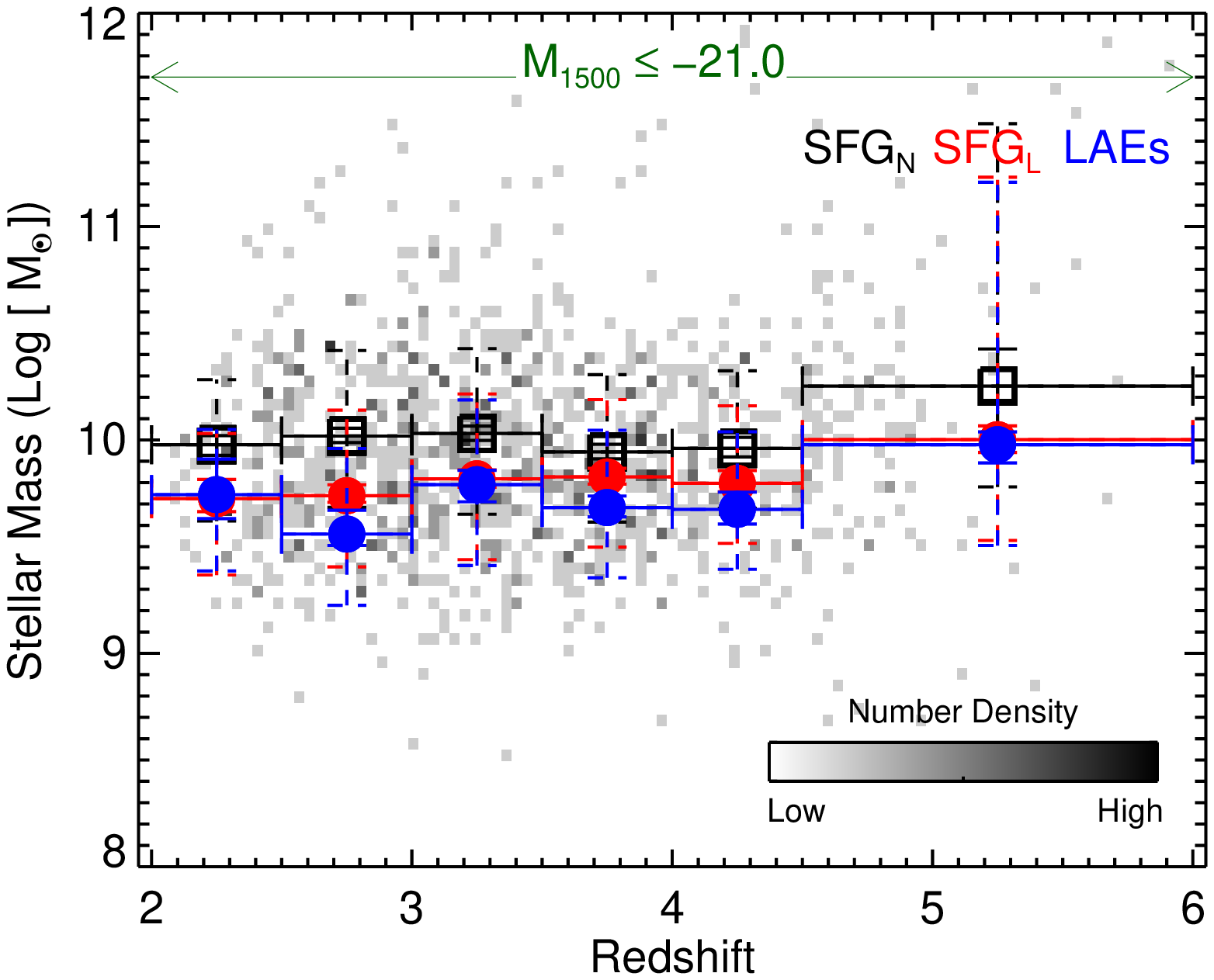}}
{\includegraphics[width=0.32\textwidth]{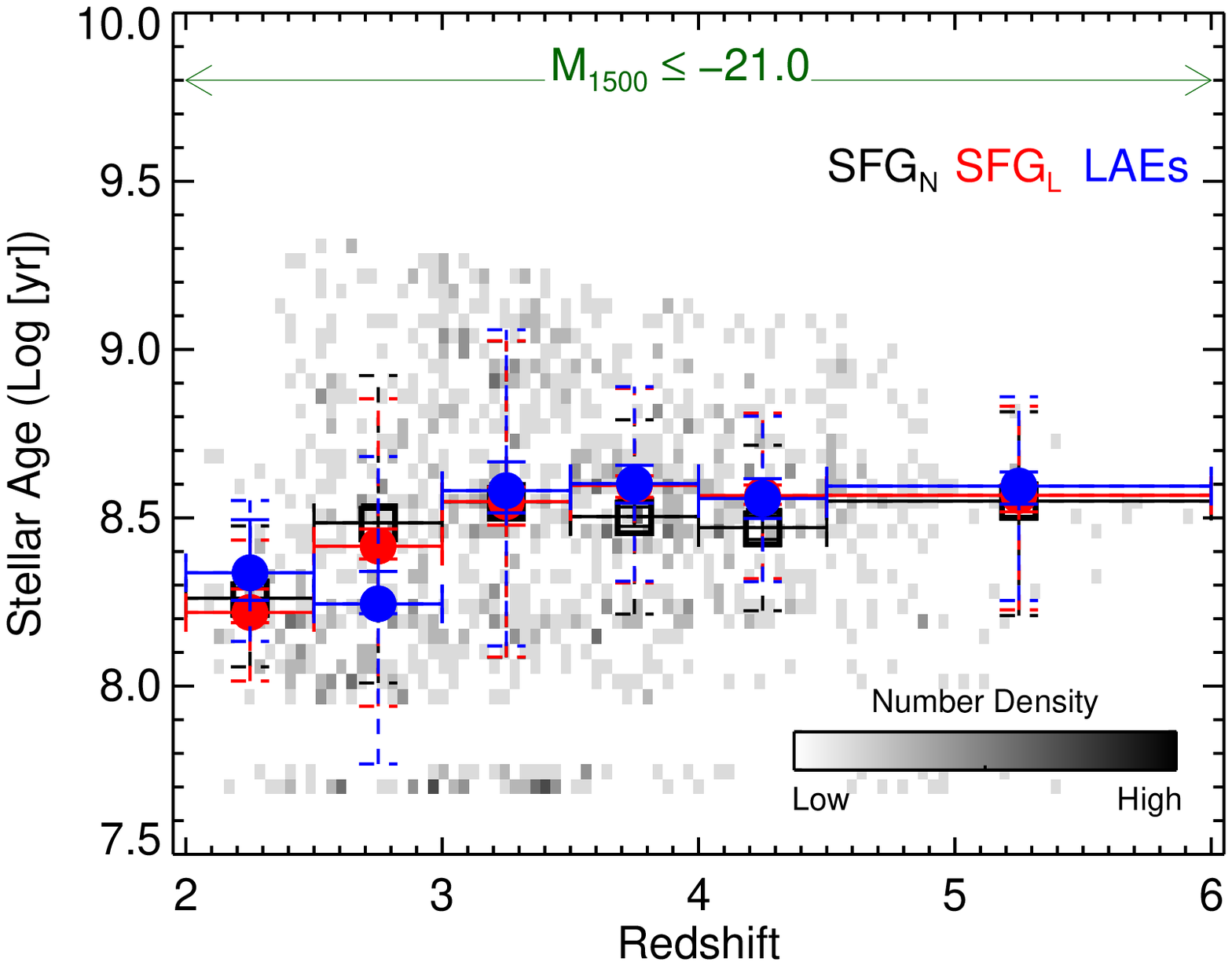} 
\includegraphics[width=0.32\textwidth]{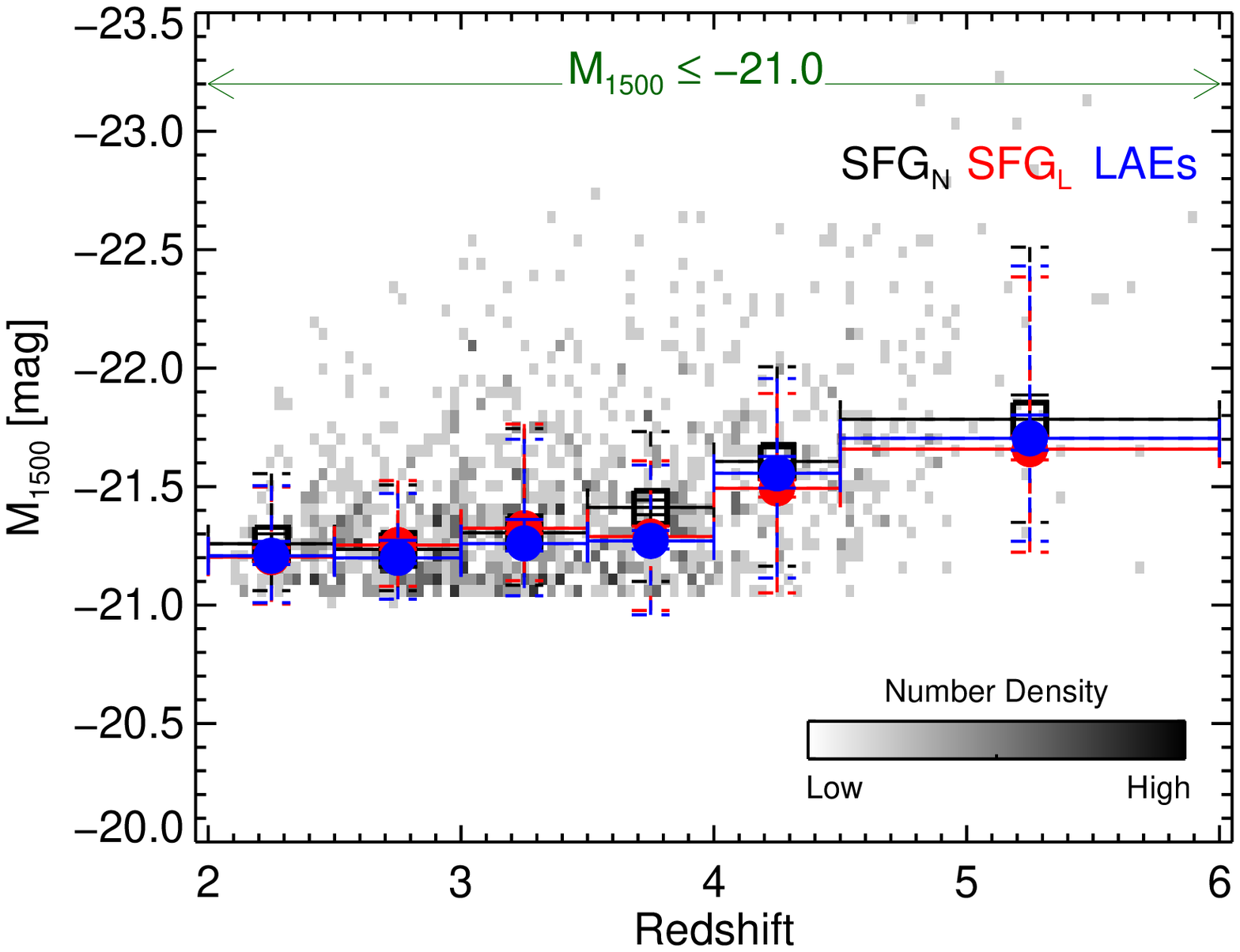}}
\caption{The best-fit SED parameters as a function of redshift for
  galaxies with M$_{\rm 1500}$\,$\le$\,--21.0. The density of points
  is color-coded as shown in the color-bar. The median values of each
  SED parameter, in each redshift bin, for the \sfgn, \sfgl, and LAE
  samples are denoted by the black squares, red circles, and blue
  circles, respectively.  The error bars in \emph{x} illustrate the
  sizes of the bins, while the errors in \emph{y} are $\pm$1$\sigma$
  scatter (dashed error bars) corresponding to the range between 16th
  and the 84th percentile values within each bin, while smaller solid
  error bars are the errors on the median values
  ($\sigma$/$\sqrt{\rm N_{\rm gal}}$). It is important to note that
  these trends in SED parameters, between LAEs and non-LAEs, are valid
  for the whole sample.}
   \label{results}
\end{center}
\end{figure}

\vspace{-0.15in}
\section{Results}
We use a large ($\sim$4000) spectroscopic sample of SFGs at
$z$\,$\sim$\,2--6 from VUDS to investigate their spectral and
photometric properties.  \figref{results} shows a comparison between
SED-based stellar parameters of the LAEs, \sfgl, and \sfgn\
samples. Here, we have applied a UV absolute magnitude cut
(M$_{\rm 1500}$\,$\le$\,--21.0), which is around M$^*$ for galaxies at
$z$\,$\sim$\,3--6 \citep{fink15}, to investigate any evolutionary
trend as a function of redshift.  The SED-based dust indicator,
E$_{\rm s}$(B-V), shows smaller values for LAEs compared to \sfgn\
galaxies. The difference between median E$_{\rm s}$(B-V) values for
LAEs and non-LAEs seems to increase as redshift increases (0.05 at
$z$\,$\sim$\,2 to 0.1 at $z$\,$\sim$\,3.5 to 0.2 at
$z$\,$\sim$\,5). This could be one of the reasons why we observe an
increase in the LAE fraction as a function of redshift
(\figref{sample}).  The SED-based SFRs depend on the amount of dust in
galaxies and hence, show small decrease in their median values for
LAEs compared to non-LAEs. This difference, on average, is
$\lesssim$0.3 dex. The SED-based stellar masses ($\lesssim$0.2 dex),
stellar ages ($\lesssim$0.1 dex), and M$_{\rm 1500}$ ($\lesssim$0.1
mag) show, on average, smaller difference between LAEs and non-LAEs,
as shown in \figref{results}. These trends in SED parameters, between
LAEs and non-LAEs, remains the same irrespective of the M$_{\rm 1500}$
cut.

The small but significant differences that we observe in the SED-based
parameters of LAEs and non-LAEs could be a direct consequence of our
sample selection. Our sample of LAEs (and non-LAEs) is selected based
on their continuum magnitudes compared to NB/emission line selection
of LAEs. The NB/emission line selected LAEs are physically different
as they probe lower continuum luminosities and typically extend to
higher EW ($\gtrsim$\,200\AA) LAEs. From various studies on this topic,
including this one, it is imperative
to note that LAEs have a wide range of stellar properties depending on
their selection criteria, luminosities, and stellar masses. A small
difference in SED-based stellar parameters, between LAEs and non-LAEs,
points to the fact that the escape of Ly$\alpha$ emission from
galaxies is a complex process, and could be affected by intrinsic
properties of these galaxies which includes, the dust
content/geometry, morphology, kinematics and interstellar medium
geometry, and/or change in the Lyman continuum escape fraction.  Our future
studies will continue to explore various aspects affecting \lya\
emission from galaxies to better understand the difference in physical
properties of LAEs and non-LAEs.

\end{document}